\documentclass{iau}
 \usepackage{graphicx}

\title[IAUS291.~~On Fermi observations of magnetars] 
{What can {\it Fermi} tell us about magnetars?} 

\author[Tong \& Xu]  
{H. Tong$^1$
 \and R. X. Xu$^2$}

\affiliation{$^1$Xinjiang Astronomical Observatory, Chinese Academy of Sciences,
Urumqi, Xinjiang 830011, China  Email: {\tt tonghao@xao.ac.cn} \\[\affilskip]
$^2$School of Physics and State Key Laboratory of Nuclear Physics and Technology,
Peking University, Beijing 100871, China Email: {\tt r.x.xu@pku.edu.cn}}

\pubyear{2012}
\volume{291}  
\jname{\mbox{Neutron Stars and Pulsars: Challenges and Opportunities after 80 years}}
\editors{J. van Leeuwen, ed.}
\begin{document}

\maketitle

\begin{abstract}

We have analyzed the physical implications of Fermi observations of magnetars.
Observationally, no significant detection is reported in Fermi observations of all magnetars.
Then there are conflicts between outer gap model in the case of magnetars and Fermi observations.
One possible explanation is that magnetars are wind braking instead of magnetic dipole braking.
In the wind braking scenario, magnetars are neutron stars with strong multipole field. A strong
dipole field is no longer required. A magnetism-powered pulsar wind nebula and a braking index
smaller than three are the two predictions of wind braking of magnetars. Future deeper Fermi
observations will help us make clear whether they are wind braking or magnetic dipole braking.
It will also help us to distinguish between the magnetar model and the accretion model for AXPs
and SGRs.

\end{abstract}


\firstsection 
\section{Introduction}
Anomalous X-ray pulsars (AXPs) and soft gamma-ray repeaters (SGRs) are peculiar
pulsar-like objects. They are commonly assumed to be magnetars, magnetism-powered neutron stars.
The traditional model of magnetars is that they are young neutron stars with both strong dipole field
and strong multipole field (Duncan \& Thompson 1992; Thompson et al. 2002).
The presence of strong dipole field ensures that they can also accelerate particles
to very high energy via the outer gap mechanism. Therefore, magnetar may emit high-energy gamma-rays detectable by
{\it Fermi}-LAT (Cheng \& Zhang 2001). These high-energy gamma-rays are rotation-powered in nature
(Zhang 2003). Detection of rotation-powered activities in mangetars will help bridge the gap between
magnetars and normal pulsars.

{\it Fermi}-LAT has observed the whole magnetar class. No significant detection is reported (Sasmaz Mus
\& Gogus 2010; Abdo et al. 2010). Then they are conflicts between the outer gap model in the case of magnetars
and {\it Fermi} observations (Tong, Song, \& Xu 2010, 2011). Below we will show what can {\it Fermi} tell us about
magnetars.

\section{Implications of Fermi observations of magentars}

We have analyzed the implications of {\it Fermi} observation of AXP 4U 0142+61 (Sasmaz Mus,
\& Gogus 2010; Tong, Song, \& Xu 2010).
It is shown that there are conflicts between outer gap model in the case of AXP 4U 0142+61
and {\it Fermi} observations. The fallback disk model for AXPs and SGRs can still not be ruled out.
In {\it Fermi} observations of the whole magnetar class, still no significant detection is reported.
This is consistent with out previous analysis (Abdo et al. 2010; Tong, Song, \& Xu 2011).
The upper limit of AXP 4U 0142+62 lies already
below the theoretical calculations for some parameter space (Tong, Song, \& Xu 2011). Future deeper
{\it Fermi} observations will help us to distinguish between the magnetar model and the accretion model
for AXPs and SGRs.

\section{Solutions and predictions}

There are two possible explanations to the non-detection in {\it Fermi} observations of magnetars.
One possibility is that AXPs and SGRs are accretion-powered systems. Then it is natural that
they are not high-energy gamam-ray emitters. Various observations of AXPs and SGRs can be explained naturally
in the accretion scenario (Tong \& Xu 2011).

The other possibility is: magnetars are wind braking. If magnetars are wind braking instead of mangetic
dipole braking, then their magnetosphere structure is different from that of normal pulsars.
Vacuum gaps (including outer gap) may not exist in magnetars. This may explain the non-detection
in {\it Fermi} observations of magnetars (section 4.2 in Tong, Xu, Song, \& Qiao 2012).

\begin{figure}[!htbp]
 \centering
\includegraphics{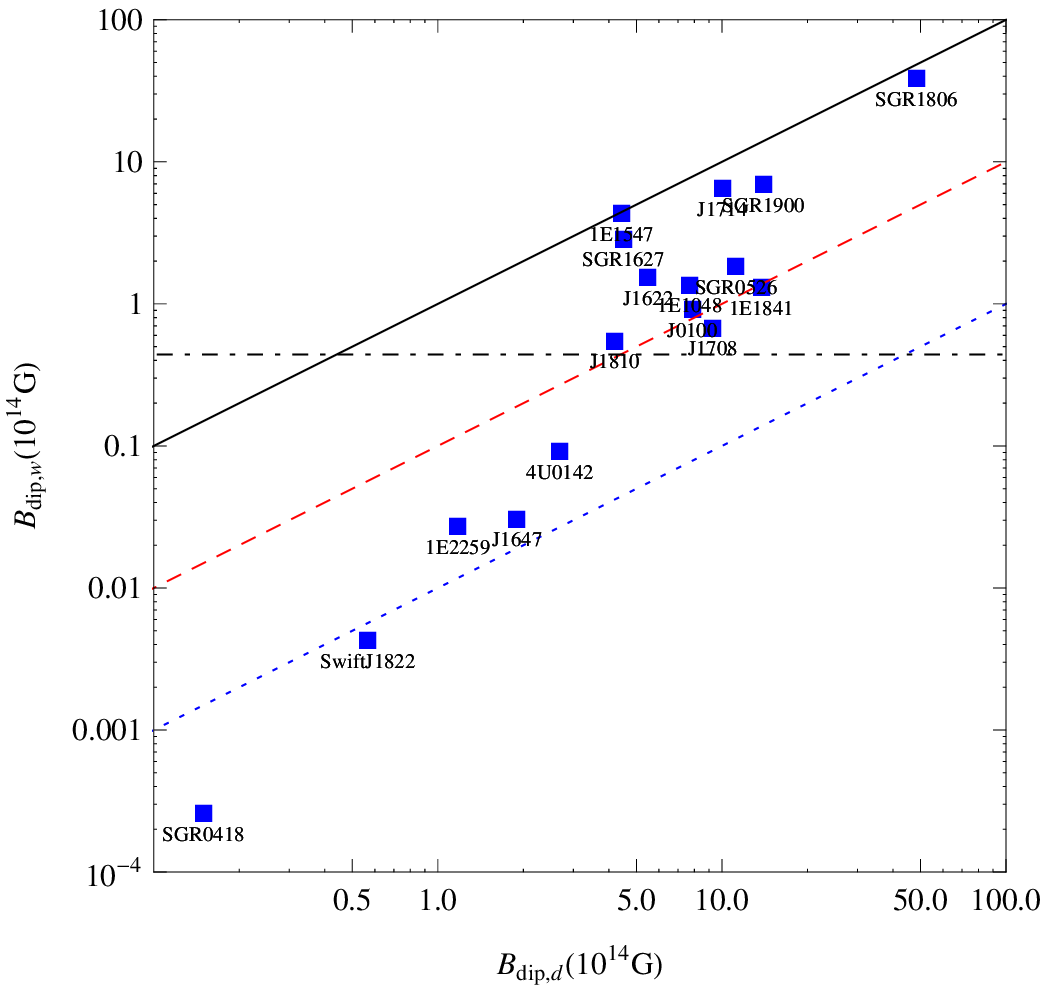}
\caption{Dipole magnetic field in the case of wind braking versus that
in the case of magnetic dipole braking. A wind luminosity $L_{\rm p}= 10^{35} \,\rm erg\, s^{-1}$ is assumed.
The solid, dashed, and dotted lines are for
$B_{\rm dip,w}=B_{\rm dip,d},\ 0.1 B_{\rm dip,d},\ 0.01 B_{\rm dip,d}$, respectively.
The dot-dashed line marks the position of quantum critical magnetic field $B_{\rm QED}= 4.4\times 10^{13} \,\rm G$.
See figure 2 and corresponding text in Tong, Xu, Song, \& Qiao (2012) for details.}
\label{fig_Bdip}
\end{figure}

In the wind braking scenario, magnetars are neutron stars with strong multipole field. A strong
dipole field is no longer required. Figure \ref{fig_Bdip} shows the correspond dipole field in the
case of wind braking and that of magnetic dipole braking

Recent challenging observations of magnetars may be explained naturally in the wind braking scenario:
(1) The supernova energies of magnetars are of normal value;
(2) The problem posed by the low-magnetic field
soft gamma-ray repeater; (3) The relation between magnetars and high magnetic field pulsars ;
(4) A decreasing period derivative during magnetar outbursts etc.
A magnetism-powered (instead of rotation-powered) pulsar wind nebula  will be one of the consequences of
wind braking. For a magnetism-powered pulsar wind nebula, we should see a correlation between
the nebula luminosity and the magnetar luminosity. The extended emission around AXP 1E 1547.0-5408
may be a magnetism-powered pulsar wind nebula.
Under the wind braking scenario, a braking index smaller than three is expected. More details are
presented in Tong, Xu, Song, \& Qiao (2012).

Considering that magnetars are wind braking (both a rotation-powered and a magnetism-powered
particle wind), many aspects of magnetars can be reinterpreted. For example, the ``low magnetic
field" magnetar SGR 0418+5729 may actually be a normal magnetar (Rea et al. 2010; Tong \& Xu 2012).
It is a little special since it has a special geometry, e.g. a small magnetic inclination angle.
Another example is: low luminosity magnetars are more likely to have radio emissions.
The reason is that low luminosity magnears may have similar magnetospheric structure to that
of normal radio pulsars (Rea et al. 2012; Liu, Tong, \& Yuan 2012).

\section{Conclusions}

{\it Fermi} observations of magnetars tell us that magnetars may be wind braking
instead of magnetic dipole braking. Future deeper {\it Fermi}-LAT observations
will help us make clear whether magnetars are wind braking or magnetic dipole braking.
It will also help us to distinguish between the magnetar model and the accretion model for AXPs and SGRs.
Therefore, deeper {\it Fermi}-LAT observations of the magnetar class in the future are highly recommended.

\section*{Acknowledgments}

This work is supported by the National Basic Research Program of China
(2009CB824800), the National Natural Science Foundation of China 
(11103021, 10935001, 10973002), and the John Templeton Foundation.


\end{document}